# A simple, static and stage mounted direct electron detector based electron backscatter diffraction system

Tianbi Zhang (0000-0002-0035-9289), T. Ben Britton* (0000-0001-5343-9365)

Department of Materials Engineering, The University of British Columbia. 309-6350 Stores Road, Vancouver, BC, V6T 1Z4 Canada

* Corresponding author: ben.britton@ubc.ca

## Abstract

To engineer the next generation of advanced materials we must understand their microstructure, and this requires microstructural characterization. This can be achieved through the collection of high contrast, data rich, and insightful microstructural maps. Electron backscatter diffraction (EBSD) has emerged as a popular tool available within the scanning electron microscope (SEM), where maps are realized through the repeat capture and analysis of Kikuchi diffraction patterns. Typical commercial EBSD systems require large and sophisticated detectors that are mounted on the side of the SEM vacuum chamber which can be limiting in terms of widespread access to the technique. In this work, we present an alternative open-hardware solution based upon a compact EBSD system with a simple, static geometry that uses an off-the-shelf direct electron detector co-mounted with a sample. This simple stage is easy to manufacture and improves our knowledge of the diffraction geometry significantly. Microscope and detector control is achieved through software application programming interface (API) integration. After pattern capture, analysis of the diffraction patterns is performed using open-source analysis within AstroEBSD. To demonstrate the potential of this set up, we present two simple EBSD experiments using line scan and mapping. We hope that the present system can inspire simpler EBSD system design for widespread access to the EBSD technique and promote the use of open-source software and hardware in the workflow of EBSD experiments.

1. **Introduction**

Electron backscatter diffraction (EBSD) is an orientation microscopy technique in the scanning electron microscope (SEM) based on capture and analysis of electron backscatter patterns (EBSPs) that contain bands of raised intensity (Kikuchi bands). Each EBSP contains information about the crystallography of the material that interacted with the primary beam and formed the pattern. By capturing multiple patterns in a grid, it is possible to generate information-rich and useful microstructure maps. The EBSD technique was first implemented in the 1970s [1] thanks to initial development of fast digital cameras [2], automated pattern indexing methods [3], as well as the implementation of computer control [4]. The experimental set up and use of the technique have gradually become more routine and standardized since the 1990s, and its popularity continues to increase.

For a technique that could be considered mature, it is remarkable that there have significant and recent developments to EBSD-based materials characterization. This has been achieved in part due to investment from commercial companies offering EBSDs systems, but also thanks to a community who have driven technique development with further iterations of software, exponential growth of computation power, and evolutions of peripheral hardware. Software developments have been driven through advances in interpretation of the patterns, which include new pattern indexing methods such as spherical indexing [5], cross correlation-based methods [6,7], and pattern matching. These developments enable careful phase classification of complex structures [8,9], high angular (misorientation) resolution EBSD analysis towards ~0.006°/$1 \times 10^{-4}$ rads [10] and opens up the possibility of measurements of relative strain, changes in lattice parameter [11,12] and the density of geometrically necessary dislocations (GNDs) [13]. These algorithms have also benefited significantly from the parallel development of improved detectors (for example [14]) and ease of access to the stored diffraction pattern data (e.g. using HDF5 formats [15]). In the hardware aspect, new and integrated experimental flows such as integrated EBSD and energy dispersive X-ray spectroscopy (EDX/S), and 3D EBSD have emerged. Particular efforts have been devoted to improving the collection efficiency, and most recently both with faster indirect systems and also with electron detection technologies for higher pattern quality, especially with direct electron detectors (DEDs) offering superior pattern resolution and signal-to-noise ratio (SNR) as compared to indirect systems [16,17] due to the direct detection and event counting mechanisms. To date, there are multiple examples where DEDs have been applied to both EBSD [14,18–20] and transmission Kikuchi diffraction (both on- and off-axis) [21,22]. In addition, DEDs can often record a stream of different types of data (e.g. events and energy) simultaneously [23], which could benefit the study of diffraction physics as well.

However, typical EBSD systems are often associated with a complex and bulky detector setup mounted from the side of a SEM instrument, which may limit the accessibility of the technique (e.g. due to the availability of a port on a particular instrument). Furthermore, in these systems the relative positions of the sample and detector are neither constrained nor well known as the sample on the SEM stage may not be accurately mounted with respect to the detector, thus there can be drift between these two positions [24]. This creates extra complications for higher resolution EBSD measurements and advanced algorithms that require precise knowledge of the projection parameters including the pattern centre (PC) [25]. To reduce the geometric uncertainties, a stage setup with a static relationship between the sample and the detector can be used. During peer review, we were made aware of an earlier "modular" EBSD system was proposed by Thaveeprungsriporn and Thong-Aram in 2000 which employed an indirect CCD camera [26], and in the available extended abstract only a single diffraction patterns was shown. We further note that similar static setups have later emerged and been applied to 3D EBSD and on-axis TKD studies in SEM [20,21,27].

In this work, we present an EBSD system based on a compact and off-the-shelf Timepix3-based DED for single spot and small area analyses using a newly designed simple and static custom-built stage for reflection-based EBSD from an inclined sample. As an extension of the prior work on modular EBSD systems, the direct electron detector further consolidates the geometry of the stage and provides higher quality

EBSD patterns due to its higher SNR. We demonstrate that co-mounting of the detector and the sample on a substage makes calibration easier to perform and maintain for an experiment. Further, we developed in-house microscope and detector control scripts based on existing application programming interfaces (APIs) to realize automated line scan and mapping with the present system and used open-source toolboxes for data processing, analysis, and visualization. Together, we hope that these simple advances further expand capabilities of the technique to even richer development opportunities and provide more laboratories access to this useful technique.

## 2. Methodology

**Detector and sample stage.** A MiniPIX Timepix3 detector (Advacam s.r.o, Czech Republic) is used in this work as the housing has a compact formfactor (80 x 21 x 14 mm$^3$) and off-the-shelf availability at (relatively) low cost. To connect the device while it is in the vacuum, a custom vacuum USB feedthrough was manufactured to connect the detector to a control computer. This is made up of two cables and a vacuum feedthrough set into a vacuum plate that fits on a port of the SEM. The DED connects to a USB mini 4 pin to LEMO 4 pin cable. This cable connects to the vacuum feedthrough via a standard 4-pin LEMO coupler. On the outside of the SEM, a LEMO 4 pin to USB Type A cable connects to the computer.

The detector is based on the Timepix3 chip [23] with a USB 2.0 readout interface [28]. This detector has been purchased with 100 μm-thick Si active layer bump-bonded to the readout chip. An aluminum shutter protecting the sensing layer is available on the detector, which is manually opened prior to experiments and closed after use. During operation this layer is biased at 50 V. The readout chip has an array of 256 x 256 read out sites with a 55 µm pitch that forms in effect a 256 x 256 pixel read out array, forming a total sensing area of approximately 14x14 mm$^2$. The energy threshold was set at the minimum level of 3.01 keV based on factory calibration. The detector operates on the frame readout and event counting mode, and raw patterns were captured using the frame readout mode in this work. The raw patterns are captured by the PIXET Pro software and its accompanying API (version 1.7.8) and stored as Hierarchical Data Format (HDF5) files.

To enable EBSD analysis, the detector and sample are co-mounted within an aluminum substage (see Figure 1). The geometry of the substage was designed to orient the detector plane such that the pixel array X-axis is aligned with the tilt axis of the SEM stage, and the pixel array Y-axis is parallel to the incoming electron beam. This substage can be mount on most SEM stages easily, as we have designed it to have a small form factor, and it mounts on the Zeiss Sigma SEM stage with a dovetail and on the TESCAN AMBER-X with a pin stub mount.

The sample is mounted on an aluminum wedge mount, which is mounted on the substage base plate. The surface sample when mounted of the wedge presents a 70° inclination angle, if the sample is well mounted. This sample wedge mount was designed to accept a 12.5 mm pin stub for sample sizes of approximately 10 x 10 mm$^2$ mounted at the centre of the stub and present it to the detector so that $PC_x$ and $PC_y$ are located within the detector, which aids EBSD analysis using conventional algorithms. The wedge mount can be moved to provide different sample-to-detector distances to provide a set of four nominal detector distances (i.e. z-component of the PC, $PC_z$) available: 1.30, 1.00, 0.80, 0.50 (as defined in fractions of the pattern width, as per Britton et al. [29]). These different mounting positions provide access to a range of angles subtended across the captured pattern from ~75° to ~124°. Note that the exact detector distance and angle subtended will also depend on sample thickness, surface roughness and the location examined on the sample.

With this set up the centre of the sample should ideally be mounted towards the centre of the stub to maximize the SNR of the patterns obtained from the examined area. However, there is reasonable flexibility in this design as the high dynamic range and high SNR of the direct detector enables a wide range of PC values to be successfully indexed and EBSD mapping can be conducted from anywhere across at least a 7 x 7 mm$^2$ area within the sample (as we show in Section 3).

The design of this substage has been optimized to provide co-mounting of the sample and the detector, as this simplifies installation of the system, removal from the chamber after use, and transfer between different instruments. Furthermore, the fixed nature of the stage greatly simplifies the geometry of the EBSD setup for the analysis as the sample and detector are physically coupled.

**Microscopy.** Microscopy was performed both within a Zeiss Sigma field-emission gun SEM (single spot and scanning) and a TESCAN AMBER-X plasma focussed ion beam field-emission gun scanning electron microscope (single spot analysis; only using the SEM column). For analyses presented in this work, all EBSD patterns were captured at 30 keV primary electron energy, 1 nA (TESCAN instrument) or 1.66 nA (Zeiss instrument) beam current, and an exposure time per frame of between 0.015-0.02 s.

This combination of probe current and exposure time limits the detector dose per unit time and avoids saturating most of the pixels of the Timepix3 detector (counts > 1022 per pixel per frame). To avoid artifacts associated with the edge of the detector pixel array, the edge pixels are cropped out from each raw EBSD pattern resulting in patterns that contain 252 x 252 usable pixels.

**Geometry and conventions.** To aid analysis, we briefly outline our conventions, which are shown in Figure 1 and share conventions with those described by Britton et al. [29]:

1. We use right-handed sets of axes.
2. The detector has three axes at right angles to each other with a subscript 'd', $(X_d, Y_d, Z_d)$.
3. $Y_d$ points upwards towards the pole piece of the microscope, and $X_d$ is aligned within the microscope to point along the tilt axis of the SEM.
4. The electron micrograph scan grid has a subscript 'e', $(X_e, Y_e)$ and is described in terms of scan point coordinates which have no unit.
5. The detector is aligned with respect to the electron micrograph scan grid such that $X_d$ and $X_e$ are parallel, but point in opposite directions.
6. $Y_e$ points down the electron micrograph, as this is how the beam is scanned to generate SEM micrographs.
7. The sample has three axes at right angles to each other with subscript 's', $(X_s, Y_s, Z_s)$.
8. $Z_s$ points out of, and is normal to, the sample surface.
9. $X_s$ and $X_d$ are parallel.

10. The angle between $Z_s$ and $Z_d$ is determined by the tilt of the sample. For a sample tilted 'up to 70°' this angle is 20°.
11. The pattern centre is described with $\text{PC} = (\text{PC}_x, \text{PC}_y, \text{PC}_z)$.
12. The pattern centre is described with respect to the top left of the pattern, so $\text{PC}_y$ points opposite to $Y_d$.

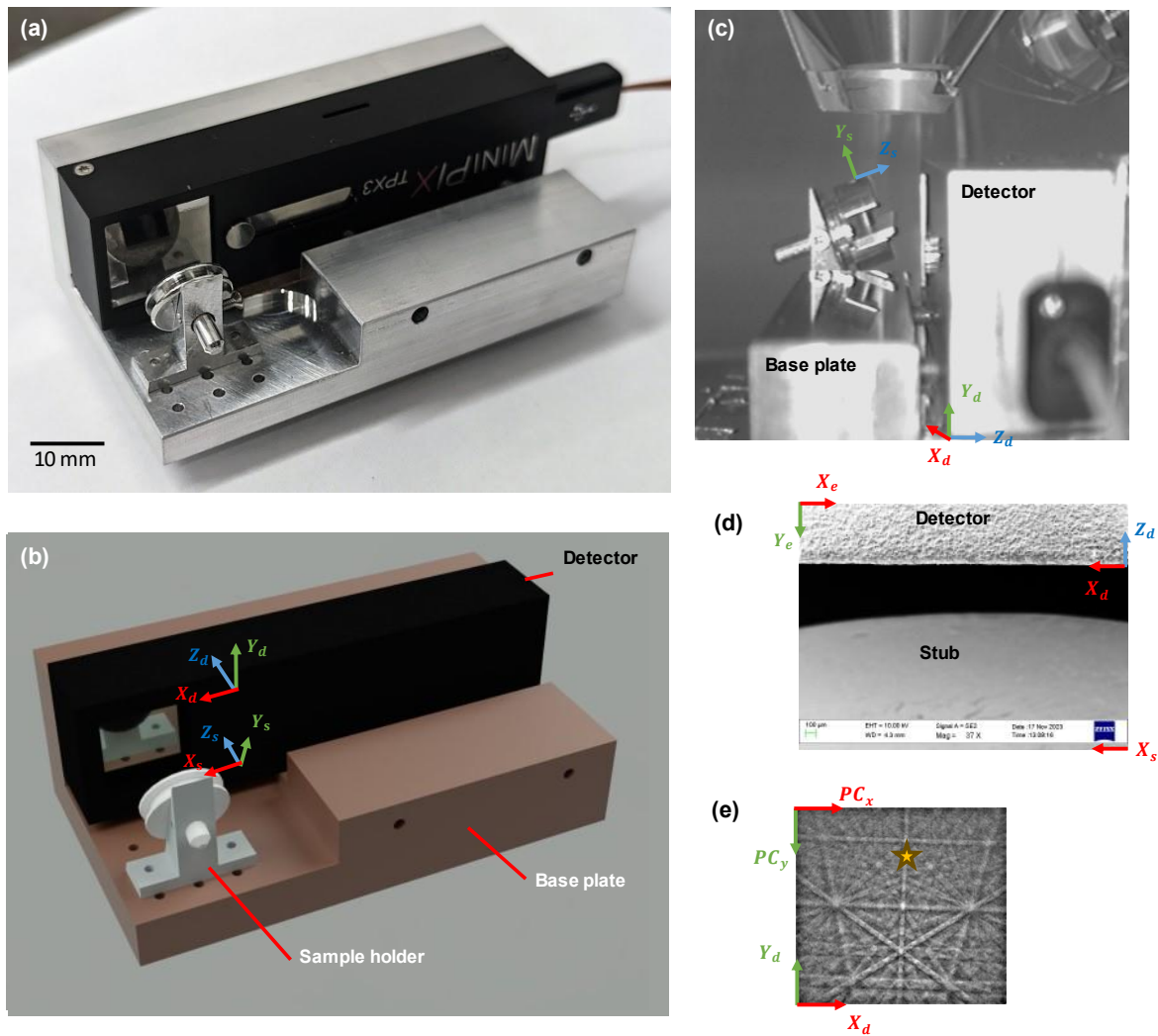

Figure 1. (a) Picture of the detector in the aluminum base stage with the sample holder installed. (b) CAD drawing of the stage assembly. The detector is in black with a reflective screen; the base plate is colored brown; and the wedge sample holder and specimen stub gray. (c) Infrared chamber scope image of the EBSD stage inside the Zeiss SEM chamber. (d) A SEM micrograph used to align the detector with respect to the electron micrograph scan grid; (e) An example EBSD pattern with the coordinate systems for the detector and pattern centre indicated. The detector axis system has subscript 'd', the sample axis system has subscript 's', the detector axis system has subscript 'd', the electron micrograph scan grid has subscript 'e'.

In practice, these coordinate systems are established by alignment of the detector and the electron micrograph grid (Figure 1e). Next the axis convention of the sample and pattern and the orientation and alignment of the stage to the SEM scan grid are verified by simply imaging the stage at low magnification. This is achieved by moving the electron beam across a single crystal sample and observing movement of the diffraction pattern features within PIXET Pro to the EBSPs as captured in real time (an example is given in Section 3). In this step, the raw pattern captured are rotated such that $X_d$ and $X_s$ are colinear.

With this set up established, individual patterns can be captured by holding the beam on a specific location within a sample using 'spot mode' within the SEM control software for the Zeiss SEM and via the 'Analysis and Measurement tool' in the TESCAN Essence software for the AMBER-X.

As is in conventional EBSD, to generate a map the electron beam is scanned across a raster grid and an EBSP is captured at each grid position. For this manuscript, this is achieved within the Zeiss SEM using a custom C# program developed based on the Zeiss SmartSEM application programming interface (API) (version 3.6) which allows user inputs of map size and step size. This scan generator implemented in the custom program uses 'spot mode' to move the beam across the SEM image, and therefore the nominal step size for the EBSD scan is set to be integer multiples of the pixel size of the SEM scan raster, and this is obtained directly from the microscope control program and saved into a text file for later use.

To capture patterns, the detector is controlled by a Python 3.7 program based on the PIXET API shipped with the Timepix3 detector, where detector properties such as exposure time and the number of frames integrated can be adjusted by the user as required. Within this Python program, the pattern capture can be synchronized with movement of the electron beam.

**Pattern Analysis.** Once the patterns have been captured and stored, offline data processing and analysis were performed with custom MATLAB scripts that are made available within AstroEBSD. Raw patterns are background corrected using a Gaussian filter, converted to 16-bit .tiff images and then analyzed by AstroEBSD using both Hough transformation-based indexing (maps and initial single pattern analysis) and pattern matching with PC refinement (single patterns) [7,30].

Once the patterns were labelled with the beam position, crystal orientation and associate pattern metrics, visualization of the resultant EBSD maps was performed by the MTEX toolbox [31]. When the crystal orientations are reported (e.g. for the mapping experiment), the orientations are rotated into the frame of the micrograph (i.e. the sample coordinate system). Note that in the mapping experiment, there is no post processing or data clean-up of the mapped data.

**Materials.** Two samples are used in this work to demonstrate capabilities of the system:

(1) a single crystal semiconductor Si(001) wafer (0.5 mm thick) to demonstrate the geometry of the experiment via single pattern captures and systematic line scans. This sample is used as it is nearly defect free and does not require sample preparation to produce high quality patterns, and the crystal is easy to align in the sample holder as the low energy cleavage surfaces of Si are the {110} faces.

(2) a polycrystal Cu sample to showcase simple polycrystalline mapping. The Cu sample was cut from a 0.2-in diameter high purity Cu rod and was first mechanically polished to 1 micron and then ion-polished by a GATAN PECS-II system at a beam energy of 5 keV and incident gun tilt of 4°.

## 3. Results

**Individual patterns.** Examples of raw and background corrected Si(001) patterns at four different sample positions (and hence camera lengths) are shown in Figure 2, as well as patterns with 1 and 2—integrated frames for sample position 1.

As the probe current used for this experiment is 1.6 nA, 1 frame captured in of 0.02s exposure results in a total incident dose on the sample of $2 \times 10^8$ electrons (the incident does on the detector is dependant on the sample and the detector position). This single frame contains sufficient Kikuchi band contrast for indexing using the Hough transform, and the higher number of integrated frames increase the SNR of the patterns.

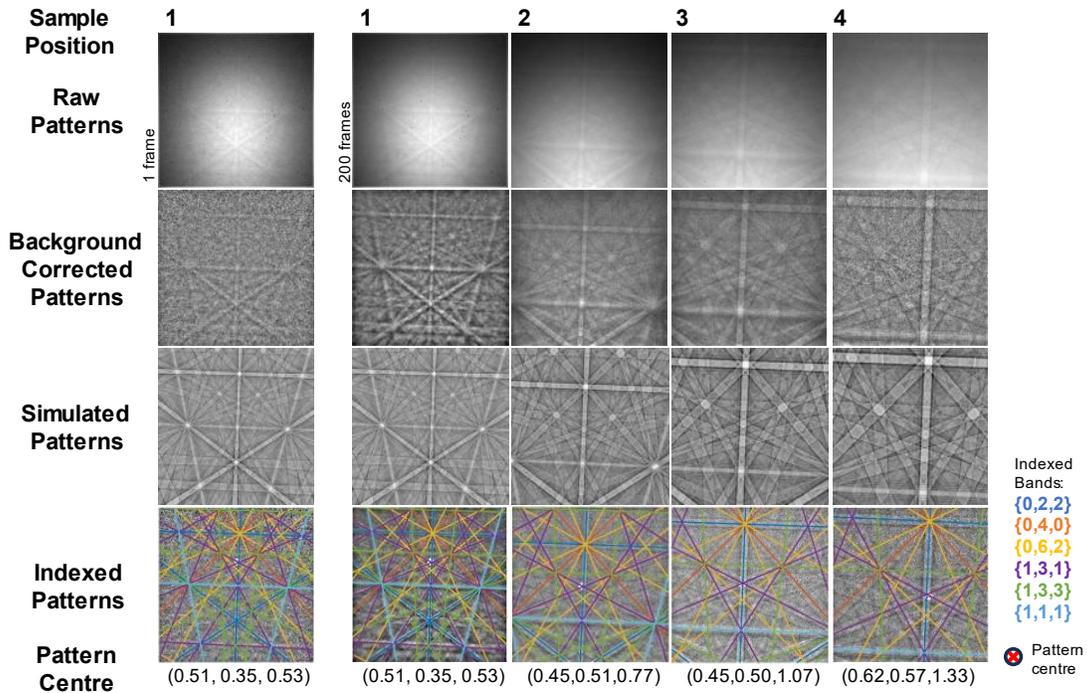

Figure 2. Raw, background corrected, simulated and indexed EBSPs of the Si(001) sample captured at 4 different sample positions, i.e. camera lengths. Color codes of indexed bands are shown (for interpretation of the color coding please see the online version). Corresponding pattern centres are listed on the top and plotted on the index patterns.

To illustrate that the static set up can be used to map large samples, Figure 3 shows EBSPs captured from the corners of the Si(001) sample. Hough based pattern centre determination within AstroEBSD reveals that the range of PC values for these patterns extend between PC = [0.260, 0.360, 0.961] and [0.787, 0.736, 1.099], and for example a $PC_x$ shift of 0.5 indicates a beam shift of 7 mm for this detector. This example shows that EBSPs captured by the detector index well for a reasonably large area, without having to move the sample with respect to the detector.

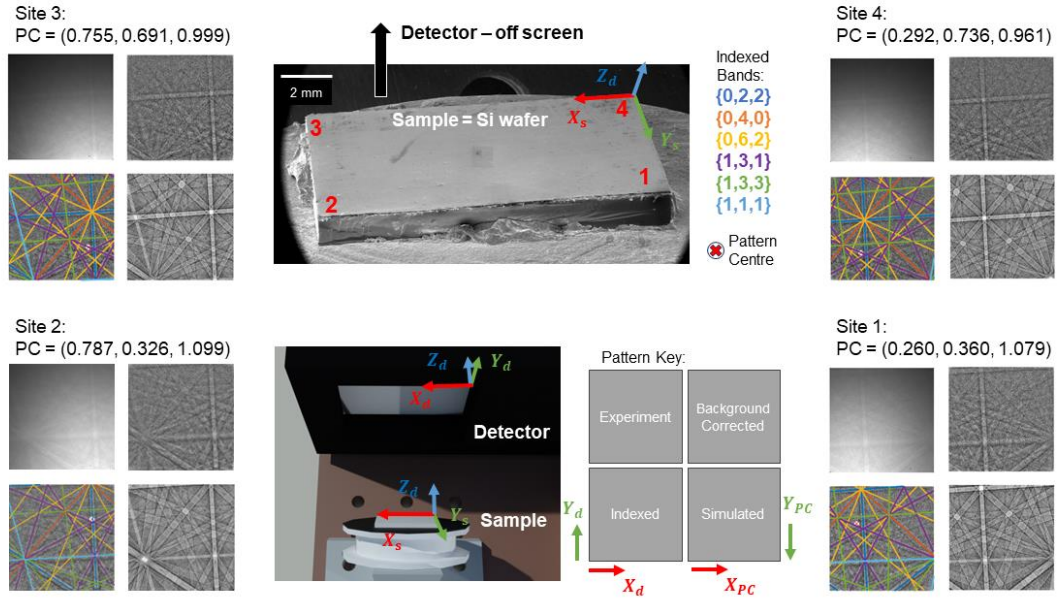

Figure 3. SEM secondary electron image of the Si(001) sample and a CAD drawing demonstrating geometry of the stub sample and the detector is shown below the SEM micrograph. Raw, background corrected, indexed and simulated EBSPs captured from four corners of the Si(001) sample are shown for each site around the SEM micrograph. Color codes of indexed bands are shown (for interpretation of the color coding please see the online version). Corresponding pattern centres are listed on the top and plotted on the index patterns.

**Line scan and system calibration.** Line scan analysis of a single crystal is used to verify and calibrate the geometry of the apparatus, and to calibrate the PC model that is essential for analysis of electron beam scan-based maps. To recap, analysis of the experimental patterns requires knowledge of the camera projection and sample geometry, specifically the PC which is described in terms of the shortest perpendicular distance between the interaction volume and the detector ($PC_z$) and the location of the coincidence between this vector and the plane of the screen ($PC_x$ and $PC_y$). In an electron beam scan, the PC of each pattern will vary depending on the geometry of sample scan grid (including sample tilt) and the detector coordinate system. During a horizontal line scan, the beam is moved along a horizontal line to a series of SEM scan grid points with a constant $Y_e$ coordinate and varying but equally spaced $X_e$ coordinates. The vertical scan can be defined in a similar manner and the geometry of this is shown in Figure 4.

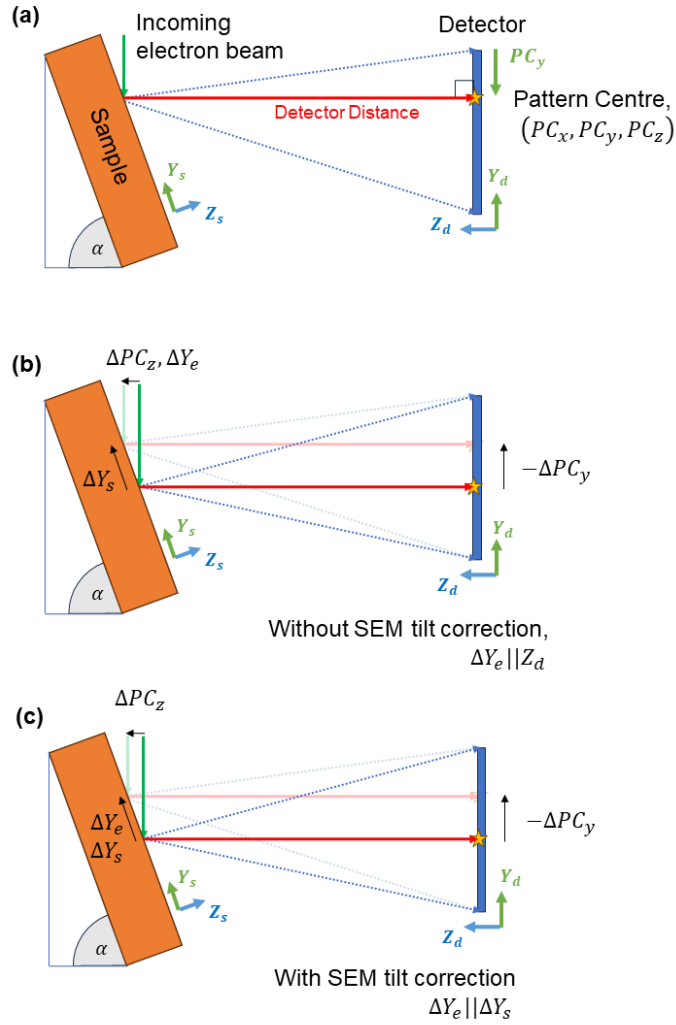

Figure 4. Illustrations of the (a) sample and detector geometry; detector reference frame, and geometries of pattern centre shift (b) without SEM tilt correction, and (c) with SEM tilt correction.

To model the pattern centre movement, we relate the beam scans $[\Delta X_e, \Delta Y_e]$ and the initial pattern centre $[PC_x, PC_y, PC_z]$. When the electron beam is moved in the scan grid by $[\Delta X_e, \Delta Y_e]$ the resultant PC shift is denoted by $[\Delta PC_x, \Delta PC_y, \Delta PC_z]$. Now we consider a line scan along $X_e$, with a SEM step size $N$ µm, a pattern height of $Q$ pixels and detector pixel pitch $M$ µm. The shift in pattern centre given by a beam shift of $\Delta X_e$ is:

$$\Delta PC_x = -\frac{N\Delta X_e * 1\text{ pixel}}{M*Q} = \frac{\Delta X_s * 1\text{ pixel}}{M*Q} \qquad 1$$

Noting that $-N\Delta X_e = \Delta X_s$. Therefore, if the magnification of the microscope is well calibrated, then we anticipate that the gradient of a variation of pattern centre with beam position will be $\frac{N}{MQ}$.

For a vertical scan (a scan along $Y_e$), as per Figure 4 let $\alpha$ be the tilt angle of the sample with respect to the detector screen. Therefore, from the geometry shown in Figure 4(b) where tilt correction has not been applied there will be a variation of pattern centre from a line scan along $Y_e$, by:

$$\Delta PC_z = \frac{N\Delta Y_e * 1\text{ pixel}}{M*Q} \qquad 2$$

And

$$\Delta PC_y = -\frac{N\Delta Y_e * 1\text{ pixel}}{M*Q}\tan(\alpha) \qquad 3$$

Therefore

$$\tan \alpha = -\frac{\Delta PC_y}{\Delta PC_z} \qquad 4$$

Similarly, for data collected when SEM tilt correction is applied, as shown in Figure 4c, then the model describing the movement of pattern centre with scanning position can be written as:

$$[\Delta PC] = [\Delta X_s, -\Delta Y_s \sin\alpha, \Delta Y_s \cos\alpha]\frac{N*1\text{ pixel}}{MQ} \qquad 5$$

To test these relationships, a horizontal and a vertical line scan at a nominal step size of 10 µm were performed with 20 measurements equally spaced on the Si(001) sample with initial $PC_z$=0.75. Curve fitting was performed within MATLAB, and for the linear fits, the PC and beam shift data were normalized based on the mean value and standard deviation of the data series and therefore the intercepts of the linear fits were forced to 0.

Figure 5 shows the fitted PC shifts for the horizontal and vertical scans as a function of beam shift distance in µm. The fitted gradients and their 95% confidence intervals are reported in Table 1. From the vertical line scan, the tilt angle $\alpha$ is estimated to be 66.25°±0.32° (95% confidence interval). This deviation from the intended 70° tilt is likely because of inaccurate mounting of the stub sample (this can be seen within careful analysis of the image shown in Figure 1). From the horizontal scan, the step size is estimated to be 9.766±0.006 µm (errors given by 95% confidence interval), which is within 2.8% of the expected SEM step size 10.05 µm as reported by the microscope, and proper application of ASTM E766-14(2019) [32] results in precise calibration of the microscope magnification to within 5% or better.

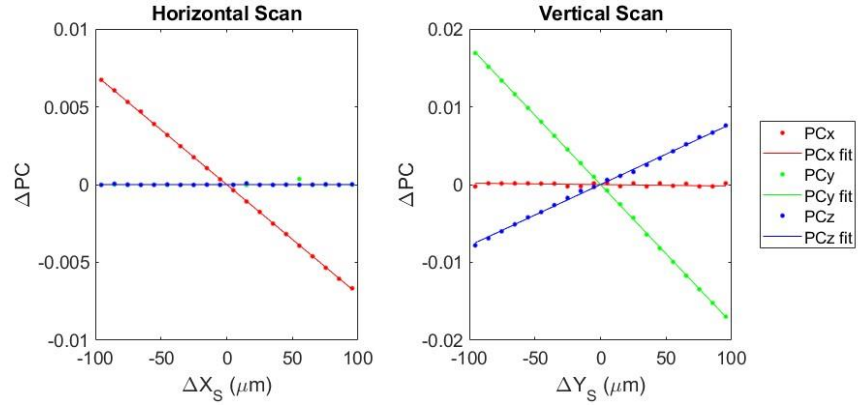

Figure 5. Pattern centre shift as a function of beam shift for horizontal and vertical line scans on the Si(001) sample.

|  | Slope of $\Delta PC_x$ vs $\Delta X_e$ | Scan Step Size (µm) | Slope of $\Delta PC_y$ vs $\Delta Y_e$ | Slope of $\Delta PC_z$ vs $\Delta Y_e$ | Tilt Angle (°) |
|---|---|---|---|---|---|
| Mean value | 0.00071 × $10^{-4}$ | 9.766 | -0.0018 | 0.00079 | 66.25 |
| Uncertainty (95% CI) | $\pm 2.6 \times 10^{-6}$ | ±0.006 | $\pm 8.0 \times 10^{-6}$ | $\pm 2.6 \times 10^{-6}$ | ±0.32 |

Table 1. Calibration values obtained from the line scan experiments

**Microstructure mapping.** For mapping of the Cu sample, a 20 x 20 point map was captured using tilt correction (assuming 70° tilt) in the microscope software with a step size of 10 µm. The sample wedge was mounted in position 1, and $PC_z$ was measured as 0.59 for the first pattern.

Analysis of the patterns was performed using a MATLAB script and AstroEBSD routines, where a PC refinement algorithm was applied for the first pass of Hough based indexing for every point in the map. Based on Equation 6, the variation in pattern centre across the map was then fit to a surface using the curve fitting toolbox within MATLAB, assuming that the pattern centre positions each varies as a function of beam shift in X and Y:

$$PC_i = a + b\, \Delta X_s + c\, \Delta Y_s$$

where $i = x, y, z$    7

The fit was performed using robust statistics in MATLAB, with an iterative reweighted least squares algorithm. The resultant fitting parameters are listed in Table 2.

| | | $a$ | $b$ | $c$ |
|---|---|---|---|---|
| $PC_x$ | Fitted value | 0.5250 | $7.58 \times 10^{-5}$ | $-1.01 \times 10^{-6}$ |
| | Uncertainty (95% CI) | ±0.0001 | $\pm 2.0 \times 10^{-6}$ | $\pm 2.00 \times 10^{-6}$ |
| | | | | |
| $PC_y$ | Fitted value | 0.4518 | $1.39 \times 10^{-6}$ | $6.05 \times 10^{-5}$ |
| | Uncertainty (95% CI) | ±0.0001 | $\pm 2.22 \times 10^{-5}$ | $\pm 2.20 \times 10^{-6}$ |
| | | | | |
| $PC_z$ | Fitted value | 0.6017 | $-9.41 \times 10^{-7}$ | $2.91 \times 10^{-5}$ |
| | Uncertainty (95% CI) | ±0.0001 | $\pm 1.75 \times 10^{-6}$ | $\pm 1.80 \times 10^{-6}$ |

Table 2. Fitting parameters of pattern centre component as a function of beam shift distance ($PC_i = a + b\, \Delta X_s + c\, \Delta Y_s$).

Using similar relationships to the line scan data, the tilt angle of the sample was found to be 64.32° ± 2.22° and the scan step size 10.46 ± 0.28 µm (again within 5% the expected value). These measurement uncertainties are reported as the 95% confidence interval, as determined by the fitted model.

After this PC model was determined, the model was used to fix the pattern centre position for each point in the map and then a second Hough-based index was used to determine the crystal orientation across the map. Figure 6 shows the mapped region together with the resultant crystal orientation map and example patterns (as captured, dynamical simulations, and the indexed experimental patterns). Note that since the actual sample tilt was 64.32° while the microscope tilt correction was wrongly applied at 70°. To correct for this effect on the microstructure map, a correction term of $\frac{\cos 64.32°}{\cos 70°} =$ 1.27 was applied to the spacing of Y scan grids. It was also found that using the new PC model, the overall mean angular error has been reduced from 0.0096° with a standard deviation of 0.0313° to 0.0045° with a standard deviation of 0.0089°.

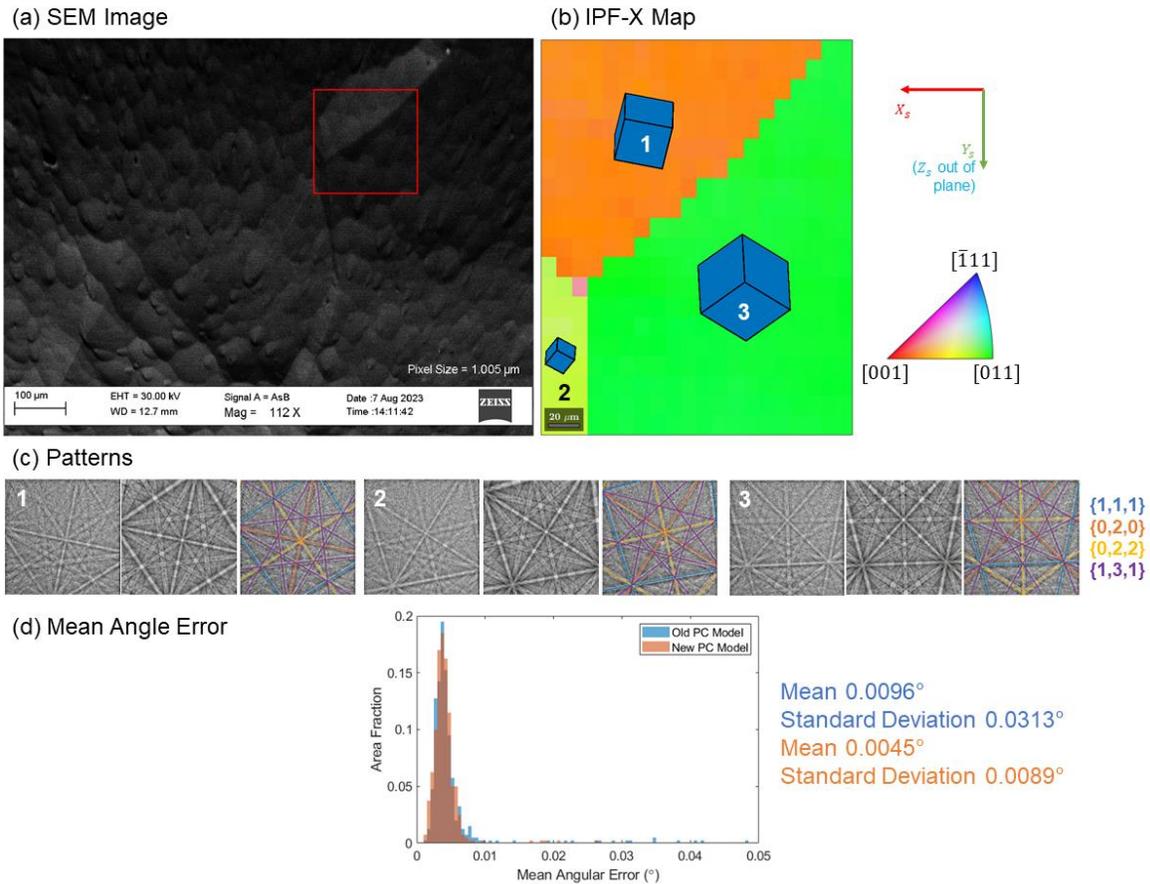

Figure 6. (a) SEM backscattered electron image of the Cu sample (EBSD mapped region in red box); (b) crystal orientation map (IPF-X colored with overlaid cubes to illustrate grain orientations) after correction of Y coordinates; (c) representative background corrected EBSPs from the three major grains identified on (b) and corresponding captured, dynamically simulated, and indexed EBSPs. Color codes of the indexed bands are shown on bottom right; (d) Distribution of mean angular errors of the map indexing using the old and the new PC models (for interpretation of the color coding please see the online version).

## 4. Discussion

We have presented a simple EBSD system using a compact and static geometry, with a DED mounted on the SEM stage. This system, when combined with open-source EBSD pattern and orientation analysis algorithms, can enable the capture and analysis of EBSD patterns across relatively large samples, as demonstrated here with points indexed from a Si sample across an area that spans 7 x 7 mm$^2$ (Figure 3). The static geometry, where the sample and detector are co-mounted on the same substage and fixed with respect to each other, has an advantage that the geometry of the system can be well calibrated for a sample as shown using the Si single crystal experiments here.

Obtaining patterns from the large Si single crystal shows that the static configuration produces reasonable and indexable diffraction patterns, even when the pattern centre is not optimized. In general, users should be familiar with the shape of the interaction volume used to form these patterns and other electron microscope related distortions. The shape of the interaction volume and the associated spatial distortions would guide a user as to whether they wish to conduct small scans across a sample to create a montage-based analysis, or to manage the sub-optimal electron probe shape and large area distortions which are commonly seen at low magnification in the SEM [33].

We demonstrated that the static geometry enables calibration of the stage itself as well as the scan step size of the microscope through line scans, and the fitting routes can also be applied to other scan geometries such as 2D mapping. The precision of these measurements, afforded from the high dynamic range patterns captured using this experiment, is useful as it enables measurement of the misalignment of the set up. This misalignment is readily apparent as it is integral to our open-source based analysis of the EBSD patterns. Notably from the line scan and mapping experiments is the deviation of the tilt angle from the desired 70°. Errors due to misalignment of the geometry are likely a common but underreported and underappreciated challenge for all EBSD experiments. For example, Nolze et al. presented a method to improve the precision of the absolution orientation of a sample in light of inaccuracies in the sample mounting and pattern analysis [34] and there are other distortions possible depending on the SEM scan generator and sample mounting [33]. Further improvements can be done in the design of the wedge mount to accept pin stubs, verification of the parallel nature of the top and bottom of the sample, and use of a calibrant sample (e.g. a co-mounted Si crystal) to align the $X_e$ and $X_d$ prior to pattern capturing, in order to reduce the misalignment. The variations in the tilt of the sample surface from fitting the pattern centre model are not unreasonable (Si @ 66°, Cu @ 64°) and the subsequent examination of the infrared chamber scope images revealed that the likely reason is related to the design and manufacture of the wedge and pin stubs in particular, which could be refined in future designs.

These errors and uncertainties of indexing caused by misalignments and inaccurate assumptions of the experimental geometry can be reduced and accounted for as required, as we demonstrate through the calibration steps used here via establishment of a model that varies the PC systematically with beam position but more flexibly than most models used in commercial EBSD systems.

Furthermore, if we have precise knowledge of the pixel size in this detector then we can use a linear shift along $X_e$ as a method to determine the SEM micrograph magnification as the results reported here are consistent with the expected precision as determined using ASTM E766-14(2019) [32].

It should be noted that a small number of data points (400) were used in this work to generate the 2D PC model for the Cu map, albeit a good fit. For better accuracy of the model, it is suggested that more data points (patterns) over a larger area be collected and analyzed.

This simple system uses the API control for the Zeiss Sigma SEM and access to this API can be prohibitively expensive and complicated to set up in the general case, as many microscope controls are locked up software APIs which may not be open or written in the same language as the camera control software. For follow-on work, we note that alternative scan generators could be used [35], and depending on the frame rate required even direct computer control of the microscope using mouse and keyboard automation can provide inexpensive semi-automated beam movement synchronized with pattern capture. Concurrently, we urge SEM manufacturers to open up access to the SEM to enable more transparent and more standardized software-based SEM control.

Overall, the results here show significant promise and potential of this low-cost system, especially that indexable patterns can be obtained with a short exposure time (0.02 s) with no frame integration required, similar to prior results of EBSPs captured with DEDs [36]. Specifically, the intersection of new hardware designs and open-source codes to realize a flexible and (relatively) inexpensive EBSD system in the lab.

The present stage adapted a conventional EBSD geometry, but we also note that by redesigning the substage and sample holders, similar systems can be used for e.g. tilt-free EBSD [37,38], off-axis TKD-in-SEM [39] and on-axis TKD-in-SEM [21], as well as EBSD combined with *in situ* mechanical testing (e.g. [40]), where sample movement with respect to the detector is not required. For more specialized EBSD experiments, some limitations of the compact geometry as well as the detector arise. For example, in *in situ* experiments involving elevated or cryogenic temperatures, the detector's proximity to the sample may cause it to be rapidly overheated or undercooled to unserviceable temperatures. 3D EBSD experiments can be, in principle, conducted on the present system. However, the user must be aware of damage to the sensing layer by scattered ions and contamination during the sputtering process. Though the introduction of an automated shutter to the detector may help mitigate the latter issues (for example in ref. [20]).

While the present work shows promise, we also note that the MiniPIX Timepix3 detector introduces a few limitations for this demonstration system. First, a particular disadvantage compared to conventional EBSD systems is the read out speed that restricts collection of patterns to a maximum speed of ~16 frames/s (slower when applying frame integration) due to the use of a USB 2.0 readout system and Timepix3 architecture. This could be overcome through an alternative read out system (e.g. USB 3.0) which allows for faster data transfer, but this may add complications to the feedthrough requirements as USB 2.0 requires two power and two data pins, USB 3 requires two power and 8 other pins. Second, due to the high power consumption of the Timepix3 chip heating of the detector in vacuum poses a limit on map size or the number of patterns the system can capture in a single experiment (as the detector will automatically disconnect when heated past 53 °C), raises the noise floor that reduces the SNR of the detector, and affects accuracy of energy-based (time-over-threshold) diffraction physics studies [41]. In practice, this could be overcome via a cooling system, and thermal management strategies including conducting the heat to the walls of the SEM chamber using Cu braids or active cooling with a Peltier semiconductor cooler and water-cooling loops.

The Timepix3 chip or Timepix chips in general are often compared with Medipix chips, especially the Medipix3 chip [42], due to their similarity (as the Timepix chip was based on Medipix2 [43]). The present work was motivated in part as we have a MiniPIX Timepix3 detector in our lab and have been using this for physics-based studies of the electron-matter interactions in EBSD-related experiments (not reported here). We are aware that the strength of the Timepix3 chip family lies in time-resolved measurements at low electron doses for event-based counting, while the Medipix family of chips are likely more suitable solution for higher dose, frame-based electron imaging (capturing scattered electrons which form the EBSPs) as they are configured with functions to optimize capture of 2D images, e.g. the charge summing mode, which enables correction of charge sharing among neighbouring pixels and improves the resolution of the captured electron image (EBSPs in this case). In practice, we could imagine a very similar set up to the system proposed here where the Timepix3 detector is replaced by a Medipix3 detector contained within a small form factor read out.

One aspect of the Medipix and Timepix family that is useful in this work is that as a family of hybrid pixel detectors, both Medipix and Timepix chips employ relatively large pixel pitch (55 µm) to accommodate a larger number of incident electrons for a superior single-frame dynamic range, but as a trade-off it also ultimately limits the spatial resolution of the detector within each collected diffraction pattern image. A large pixel size typically has a reduced the cost to fabricate the application-specific integrated circuit (ASIC) including charge collection and signal processing array. In practical terms it is also enables us to place the detector further from the sample to achieve a wide pattern capture angle. For example, the detector is 7 mm from the sample in stage position 1 (Figure 1) and the patterns still have a wide capture angle, with the horizontal subtended angle of 90° and a vertical subtended angle of 95°. Furthermore, a smaller number of pixels can be useful to achieve higher frame read out rates as a 256 x 256 pixel detector can achieve 16 times faster frame rates than a 1024 x 1024 pixel detector for the same bandwidth. An alternative DED technology is monolithic active pixel sensor (MAPS) which typically employs small pixel pitches (e.g. 6.5-20 µm) and larger amounts of pixels, but at a cost, as the electron dose each pixel can accommodate is lower and each frame captured is more sparse [14,17]. Thus, a higher number of frames integrated and/or more sophisticated post-processing of the patterns may be necessary. Furthermore, the mounting of the MAPS system may have a larger form factor than the system presented here and could impact the simplicity of the experimental set up and the flexibility for transferring between microscopes.

As these technologies mature, we anticipate access to newer generations of compact direct electron detectors, especially hybrid pixel array detectors with higher frame rates and better thermal management will further advance the capabilities of the present EBSD system.

## 5. Conclusions

In this work, we presented a low-cost, EBSD system with a simple and well-known geometry based on a Timepix3 direct electron detector with a reduced footprint compared to commercial EBSD systems. Using in-house microscope and detector control programs, we have demonstrated that the system is capable of performing conventional EBSD data acquisition, such as line scans and mapping experiments. Our open-source analysis approach enables more precise measurement of the sample tilt and alignment, which also enables calibration of the stage geometry by tracking the variation of PC and beam shift. The stage calibration results can then be used to enhance indexing results by reducing errors due to stage and sample misalignments. Ultimately we hope that the present system inspires the development of more open-source EBSD systems, and that simpler EBSD experimental flows continue to enable EBSD to develop as a field of microstructural characterization within the SEM.

## CRediT Authorship Contribution Statement

**Tianbi Zhang**: conceptualization, investigation, methodology, data curation, formal analysis, software, visualization, validation, writing – original draft; **T. Ben Britton**:

conceptualization, formal analysis, software, funding acquisition, project administration, resources, supervision, visualization, writing – review & editing.

## Acknowledgments

We acknowledge the support of the Natural Sciences and Engineering Research Council of Canada (NSERC) [Discovery grant: RGPIN-2022-04762, 'Advances in Data Driven Quantitative Materials Characterisation']. We would like to thank Dr. Kirsty Paton (Paul Scherrer Institut), Dr. Ruth Birch (University of British Columbia) and Dr. Alex Foden (Imperial College London) for helpful discussions; Dr. Guangrui Xia and Mr. Wes Wong (University of British Columbia) for providing the Si wafer and the Cu sample respectively; and Mr. Bernhard Nimmervoll, Mr. David Torok and Mr. Liam MacLellan (University of British Columbia) for their help in designing and manufacturing the EBSD stage. We thank an anonymous referee for leading us to prior literature on modular EBSD systems using indirect detectors.

## Data Availability Statement

Raw EBSD data is available on Zenodo (https://doi.org/10.5281/zenodo.8342489). Processing scripts are available as a part of the AstroEBSD software package version 1.2.0 (https://doi.org/10.5281/zenodo.10204583). Detector control scripts are available on GitHub (https://doi.org/10.5281/zenodo.10161485). Microscope control scripts without the proprietary Zeiss SmartSEM API contents are available on GitHub (https://doi.org/10.5281/zenodo.10161483).